# A Magnetoelectric Memory Device Based on Pseudo-Magnetization

Tingting Shen [§*][1,3], Orchi Hassan[§*][2], Neil R. Dilley[3], Supriyo Datta[4], Kerem Y. Camsari[5], Joerg Appenzeller[3,4]

1. Department of Physics and Astronomy, Purdue University, West Lafayette, Indiana 47907, United States

2. Department of Electrical and Electronic Engineering, Bangladesh University of Engineering and Technology, Dhaka 1000, Bangladesh

3. Birck Nanotechnology Center, Purdue University, West Lafayette, Indiana 47907, United States

4. Elmore Family School of Electrical and Computer Engineering, Purdue University, West Lafayette, Indiana 47907, United States

5. Department of Electrical and Computer Engineering, University of California Santa Barbara, Santa Barbara, CA, 93106, United States

[§]. Authors contributed equally to the article
*E-mail: shentingphy@gmail.com; orchi@eee.buet.ac.bd

**Abstract:** We propose a new type of magnetoelectric memory device that stores magnetic easy-axis information or pseudo-magnetization, rather than a definite magnetization direction, in magnetoelectrically (ME) coupled heterostructures. Theoretically, we show how a piezoelectric/ferromagnetic (PE/FM) combination can lead to non-volatility in pseudo-magnetization exhibiting overall ferroelectric-like behavior. The pseudo-magnetization can be manipulated by extremely low voltages especially when the FM is a low-barrier nanomagnet. Using a circuit model benchmarked against experiments, we determine the switching energy, delay, switching probability and retention time of the envisioned 1T/1C memory device in terms of magnetic and circuit parameters and discuss its thermal stability in terms of a key parameter called back-voltage $v_m$ which is an electrical measure of the strain-induced magnetic field. Taking advantage of ferromagnetic resonance (FMR) measurements, we experimentally extract values for $v_m$ in CoFeB films and circular nano-magnets deposited on Pb(Mg$_{1/3}$Nb$_{2/3}$)$_{0.7}$Ti$_{0.3}$O$_3$ (PMN-PT) which agree well with the theoretical values. Our experimental findings indeed indicate the feasibility of the proposed novel device and confirm the assumed parameters in our modeling effort.

**Keywords:** Magnetoelectric effect; Pseudo-Magnetization; Nanomagnet; Ferromagnetic Resonance (FMR); Back-Voltage; Memory

## Introduction

Charge-to-spin conversion is an important topic in novel spintronics applications since it allows conventional circuit currents and voltages to be translated into magnetization information [1-7]. In recent years, voltage control of magnetism (VCM) has emerged as a promising alternative to current control of magnetism due to its potential for better energy efficiency [8-9]. The magnetoelectric (ME) effect [10-16], which can facilitate low energy dissipation in both, the magnetic WRITE and READ process [11-16] has attracted substantial research interest. Apart from a few VCM phenomena that allow a deterministic 180 degrees switching of magnetism [17-18], VCM typically results in ±90 degrees switching of magnetization or a change in the easy-axis of the magnetization, necessitating additional assist mechanisms or complex pulsing schemes [18-23] to achieve deterministic switching. For example, in a (011)-cut PMN-PT/CoFeB heterostructure, with a DC voltage applied along the [011] direction, there would be an in-plane anisotropic strain induced due to the polarization re-orientation in PMN-PT which transfers to the CoFeB film and modifies its magnetic anisotropy. For sufficient large strain, the change in magnetic anisotropy would result in a 90º rotation of the magnetic easy axis. In our previous work, we have reported that in a PE/FM heterostructure, a magnetic easy axis expressed as a pseudo-magnetization $\mu \equiv m_x^2 - m_y^2$ can be used as a novel bit state in information processing. Here $m_x$ and $m_y$ are the components of the magnetization along the x- and y- axis, respectively, and are normalized to unity ($m_x^2+m_y^2=1$). A schematic diagram of bit state 0 and 1 in such a device is shown in Table 1 of Ref [14].

As an application of the proposed ME structure in Ref [14], we envision a 1 transistor / 1 capacitor (1T/1C) memory architecture. Our modeling results indicate that such a combination can lead to a non-volatile controllable pseudo-magnetization $\mu$ [14,16] due to the interaction of magnetism and piezoelectricity. Through an equivalent circuit model [16] that has been benchmarked against experiments [15], we show that the pseudo-magnetization can be switched between two deterministic states (WRITE) and that it can be read out through the inverse effect (READ). Note that in most proposed structures, only the WRITE operation is performed using the ME effect, while READ operation is performed using MTJ structures [11-13, 24], and only very few studies have focused on the READ through the inverse effect [14-16]. According to our theory and numerical simulations, the pseudo-magnetization can be manipulated by extremely low voltages, especially when the FM is designed as a low-barrier nano-magnet. The device exhibits ferroelectric-like behavior and can in principle operate at sub-ns speeds [16, 25], requiring <fJ of energy. The non-volatility of pseudo-magnetization in such structures allows achieving years of retention time for experimentally demonstrated magnetic and circuit parameters.

A key parameter that determines the performance of the proposed memory architecture is the so-called back-voltage constant $v_m$ that characterizes the coupling strength between the PE and FM layers. Experimentally, we have verified the WRITE operation in the above device concept and characterized $v_m$ values using Ferromagnetic Resonance (FMR) measurements in CoFeB/(011)-cut PMN-PT heterostructures as discussed below.

By applying a DC voltage across the PMN-PT substrate, an anisotropic in-plane strain is induced, which transfers to the magnetic material. When the strain is large enough, the magnetic easy axis of the CoFeB layer rotates by ±90 degrees, which is the magnetic WRITE operation [11, 14]. In order to extract the voltage induced effective field using FMR measurements, we have derived a modified Kittel formula that includes a voltage induced stress term in the LLG equation. By first fitting experimental FMR data obtained in CoFeB thin films deposited on PMN-PT substrates with the modified formula, we have successfully extracted a strain-induced magnetic field $H_s$ that modifies the magnetic easy-axis anisotropy. Moreover, the back-voltage constant $v_m$ that characterizes the coupling strength between the PE and FM layers [14, 16] has also been determined. Note that the extracted $v_m$ values using FMR are consistent with theoretical calculations, further validating our modified Kittel formula. Moreover, the experimentally determined back-voltage constant of several tens of mV is sufficiently large to support the 1T/1C memory concept proposed here.

Last, since it is understood that the actual device design will involve patterned disk-like structures, we have also studied the FMR response in circular nano-magnet arrays. Both quantities, i.e. $H_S$ and $v_m$, were found to be very similar to the ones obtained from CoFeB films. This finding is particularly relevant in that it i) highlights the validity of our modified Kittel formula irrespective of boundary conditions introduced by means of patterning (at least down to features of 200nm) and ii) that a sufficiently large response of the coupled PE/FM system is achievable for the desired device design of the 1T/1C memory cell.

**Theoretical Discussion**

The easy-axis information (or pseudo-magnetization) itself can be a state variable that can be switched between two deterministic states (WRITE) and it can be read out through the inverse effect (READ) [15]. The principle of pseudo-magnetism is general and could find use in voltage-control of magnetic anisotropy devices [19], but we focus our theoretical and experimental discussion here on ME heterostructures based on a combination of materials with piezoelectric/ ferromagnetic (PE/FM) heterostructures. One can define the energy expression associated with a PE/FM heterostructure as

$$E(Q,\mu) = \frac{1}{2C}Q^2 + Qv_m\mu - QV_{IN} - \left(\frac{E_A}{2}\right)\mu \qquad (1)$$

where $\mu$ is the *pseudo-magnetization* that defines the easy-axis for the magnet $\mu = m_x^2 - m_y^2$ (Fig.1a), $E_A = H_K M_S \text{Vol.}/2$ is the magnetic anisotropy that defines an easy-axis for the magnet where $H_K$ denotes the anisotropy field, $M_S$ denotes the saturation magnetization and Vol. denotes the volume of the magnet, $V_{IN}$ is the applied voltage and $v_m$ is the magnetoelectric (ME) back-voltage that couples the charge Q on the PE capacitor C with the pseudo-magnetization $\mu$ of the FM through the transferred strain. $Qv_m$ is the mutual coupling energy term which accounts for the piezoelectric and magnetostrictive effects in the coupled system. For simplicity, we do not explicitly add a demagnetization term to the energy that would limit the out-of-plane magnetization components and do not change the basic physics of the ME effect, however, these terms are included in our simulations. In the coupled heterostructure, $v_m$ is given by a combination of the material parameters of the PE and FM, $v_m = Bdt_{FM}/2\epsilon$ [15-16] where, B is the magnetoelastic constant of the magnet, d is the net piezoelectric coefficient of the PE layer, $\epsilon$ is its dielectric permittivity, and $t_{FM}$ is the thickness of the magnet. In our proposed device, we considered a ferroelectric relaxor, (011)-cut PMN-PT as the electroactive layer, due to its anisotropic piezo-strain resulting in a large piezoelectric constant and large electrostriction. Note that in the present context, we are interested in the piezoelectric properties of PMN-PT to create a large enough magnetoelastic response in the FM. A parallel plate capacitance of the structure is considered to be $C = \epsilon A/t$, where A is the area of the circular electrodes and t is the separation between the plates which is considered to be equal to the PE thickness $t_{PE}$, as generally $t_{PE} \gg t_{FM}$.

The fundamental operation principle can be understood from the energy equation. For a given $V_{IN}$, charge is formed on the capacitor which creates a mutual energy '$Qv_m$', which in turn induces a preferred easy axis in the magnet (x-axis or y-axis) with the magnetization lying in that axis, without a preference for a direction. Since electronic charge dynamics is typically much faster than magnetization dynamics [26], one can define the charge from the energy expression as $Q = C(V_{IN} - v_m \mu)$. Consider the case when $V_{IN} = 0$ for a low barrier magnet ($E_A \sim 0$). In this case the energy, $E(Q,\mu) = (1/2C)(Q + Cv_m\mu)^2 - (1/2)Cv_m^2\mu^2$ is minimized when $\mu = \pm 1$ and $Q = -Cv_m\mu$. So effectively, the energy barrier between the pseudo-magnetization states is $1/2Cv_m^2$. Therefore, even when $V_{IN}=0$, as long as $1/2Cv_m^2 \gg kT$ (where k is Boltzmann constant and T is temperature), $\mu$ can get spontaneously polarized and induces an internal charge $Q = -Cv_m\mu$, much like a standard ferroelectric or a ferromagnet. Self-consistent solution of the energy equation for this minimum energy condition shows this phenomenon (Fig.1b, c). The width of the pseudo-magnetization vs. voltage hysteretic loop is independent of the capacitance of the structure and depends only on the magnetoelectric voltage $v_m$ but the actual switching voltage depends on C through the anisotropy energy associated with the magnet, $E_A$. The loop is symmetric about the point $E_A/Cv_m$. It is important to note that although the charge on the load capacitor $C_L$ may leak out after writing, the pseudo-magnetization information is still preserved in the cell much like the situation in ferroelectric random-access-memories (FeRAM).

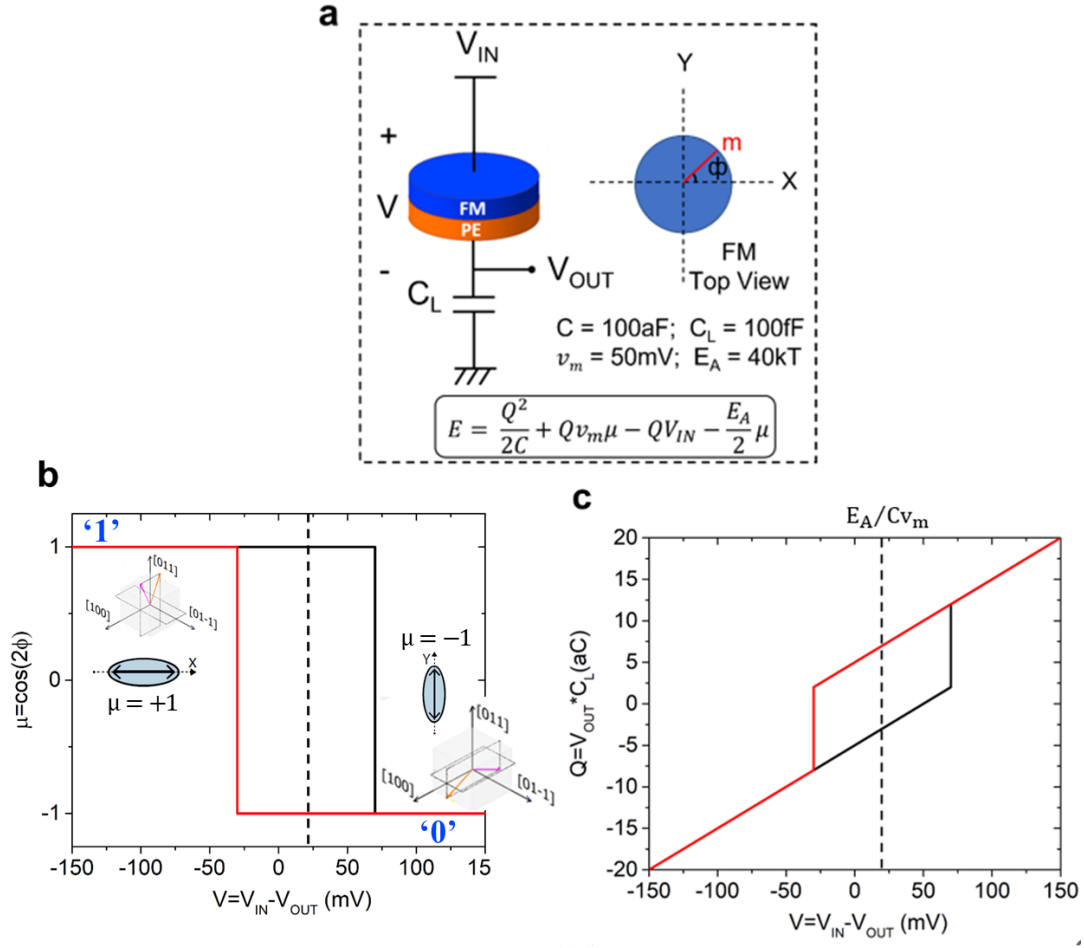

Fig.1. [Simulation] (a) Circuit schematic for the characterization of an ME heterostructure structure. Information is stored in the magnetic easy axis direction (± x or ±y) which is denoted as pseudo-magnetization, μ. (b) Change of μ due to the applied voltage V across the structure and (c) the resulting charge versus voltage characteristics in the circuit, which shows a dependence similar to conventional ferroelectrics. Both (b) and (c) are symmetric about V = $E_A/Cv_m$ (dashed vertical line). The two inset drawings in (b) show the unit cell and spontaneous ferroelectric polarizations of (001)-cut PMN-PT corresponding to '1' and '0' states. In state '1', there is a compressive strain in the y-axis, i.e. [100] direction and a tensile strain in the x-axis, i.e. [01-1] direction. In state '0', there is a tensile strain in the y-axis and a compressive strain in the x-axis. Please refer to the analysis of Fig.1 in [24] for more details.

The key parameter that defines the performance of this type of devices is a stress-induced modification to the effective field of the magnet. We define this effective magnetic field as $H_s$, which is controlled by the magnetoelectric back-voltage $v_m$ [16]. The READ-WRITE voltages and the retention time of the memory device are a direct function of this parameter. Throughout this work, we assume that the strain from the PE to the FM is completely transferred without any loss. We base this assumption on earlier theoretical calculations [16] that match experimentally observed $v_m$ values in similar material systems [15]. When comparing with experiments, this assumption may need to be carefully analyzed in specific cases [27-28].

Next, we will describe a prototypical 1T/1C memory cell that encodes pseudo-magnetization and show its READ and WRITE operation through an equivalent circuit model [16] that is benchmarked against experiments [15]. We will also show that the non-volatility in pseudo-magnetization in such structures can reach to ~years of retention time for experimentally demonstrated magnetic and circuit parameters.

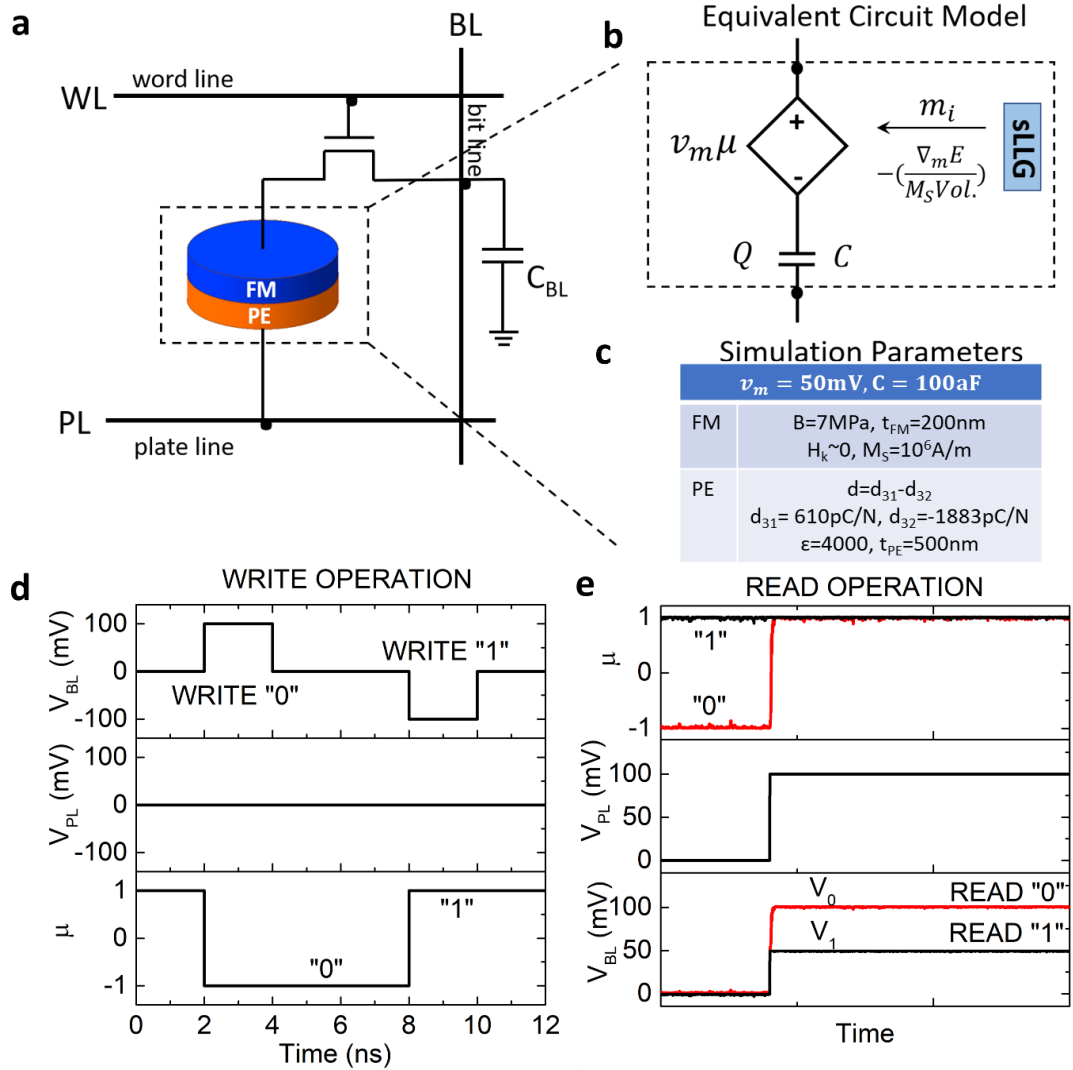

Fig.2. [Simulation] (a) Magnetoelectric 1T/1C memory cell. (b) Shows our equivalent circuit model obtained from the energy model, where the stochastic Landau-Lifshitz-Gilbert (s-LLG) equation and the circuit equations are solved self-consistently [16]. (c) Simulation parameters used for the ME circuit, based on experimental demonstrations [15]. (d) WRITE and (e) READ Operation of the memory cell mimics the scheme of FeRAM operation.

The proposed memory device combines the advantages of ferroelectric memory devices, such as energy-efficiency and high speed with those of magnetic memory such as non-volatility and high density. We envision a 1T/1C memory architecture, much like the DRAM architecture [29-30], where one end of the PE/FM capacitor is connected to the bit-line (BL) through a pass transistor and the other end is connected to a plate line (PL). The cell access is provided by the word-line (WL) as shown in Fig.2a. From SPICE simulations of the circuit model [16] (Fig.2b) we show the WRITE and READ operation of the memory cell. The simulation parameters are illustrated in Fig.2c. To write a '0' the BL is charged to '$2v_m$', PL is kept grounded and then the transistor is turned on through the WL to complete the writing process. To write a '1' a similar procedure is employed where the BL is charged to '$-2v_m$' instead. Fig.2d shows the writing process. Even after the charge on the load capacitor leaks, the internal state of the cell can be retained for a long time as long as $Cv_m^2/2kT \gg 1$. Therefore, for reading the state a read pulse needs to be applied. The BL is first pre-charged to '0V' then the access transistor is turned on which creates a capacitive divider circuit between PL and ground. When a positive read pulse is applied to PL, the voltage is divided between the ME capacitor and the bit-line capacitance ($C_{BL}$) depending on their relative values and the state of the ME device. A sense amplifier can then be used to detect these voltages. As Fig. 2e indicates, this reading process is destructive so data must be rewritten once read, similar to what is typically observed in FeRAM devices [31]. In the simulations in Fig.2, we chose parameters for the ME circuit, which are based on experimentally reported material parameters [14-15] and reasonable device dimensions. We also chose the bit-line parasitic capacitance to be equal to the PE capacitance ($C_{BL}=C$) for simplicity.

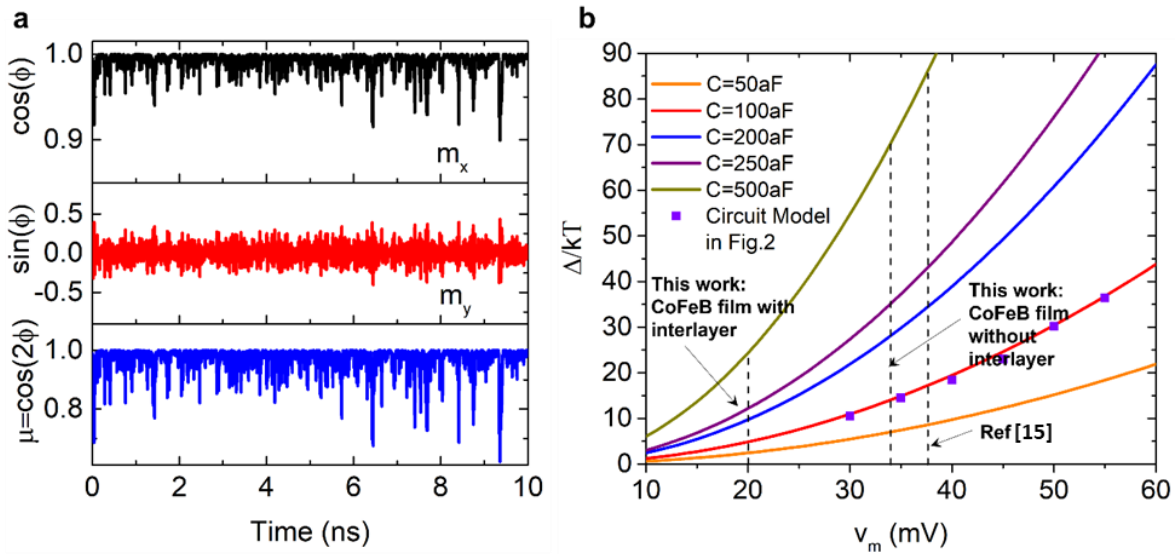

Fig.3. [Simulation] (a) The stability of pseudo-magnetization states can be measured from equilibrium fluctuations. (b) The effective stability ($\Delta$) of $\mu$ can be attributed to an effective stress anisotropy field ($H_s$) that it feels, which depends on the back-voltage $v_m$ and the capacitance value C [32]. The SPICE simulations assume operation at room temperature, T=300K and material parameters for the device structure are the same as Fig.2. The thermal barrier is extracted from 1000 samples for different $v_m$ values at C=100aF.

The energy barrier that determines the stability of pseudo-magnetization can be related to its equilibrium fluctuations (Fig.3a). By the equipartition theorem [32], the RMS value of fluctuations can be related to the energy barrier of the magnet by: $\Delta = kT/(2(1-\mu^2_{RMS}))$, where $\mu_{RMS}$ is the RMS value of the pseudo-magnetization, $\mu=m_x^2-m_y^2$. Alternatively, the equilibrium fluctuations can be obtained from the Boltzmann law [16] which yields practically identical results above $\Delta=10kT$. Fig.3b shows the extracted thermal barrier from 1000 samples, for different magnetoelectric voltages for a constant capacitance C. The results agree well with an analytically derived value of $Cv_m^2/2$. An important aspect for a non-volatile memory is the retention time $\tau_r$ and the industry standard requires $\tau_r > 10$ years for memory applications, which translates to a magnetic stability factor of >40 kT [27-28]. Figure 3b indicates that our proposed structure can achieve effective thermal stability values >40kT for experimentally demonstrated parameters of this work and others [15] for reasonable capacitance values. The information thus can be robustly retained for years until it is read-out, unlike DRAMs or FeRAMs which need periodic refreshment even if the information is not being read.

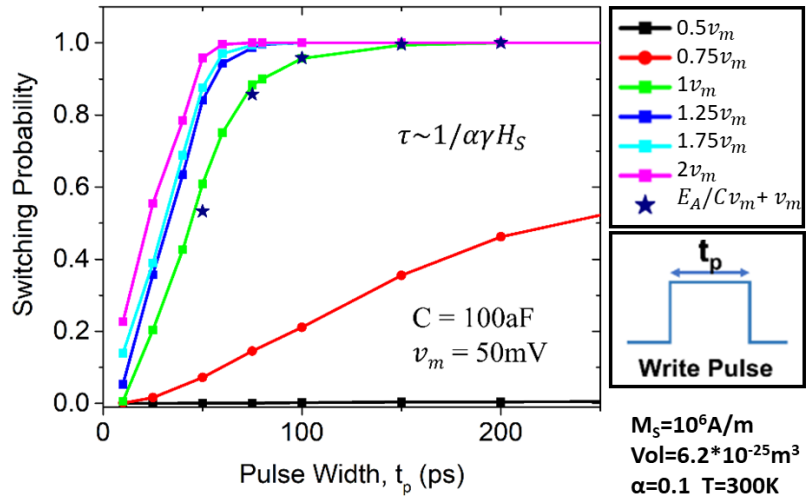

Fig.4. [Simulation] Switching probability of pseudo-magnetization is calculated from 1500 samples for different magnetoelectric back-voltage amplitudes (from $0.5v_m$ to $2v_m$) and pulse widths for the applied write pulse. The magnetic parameters used for the simulation are: $M_s = 10^6$A/m, Vol=6.2x10$^{-25}$m$^3$, and $\alpha$=0.1. The simulations are done at room temperature, T=300K. Sub-ns switching speeds ($\tau$) can be attained due to high stress fields (H$_s$=CV$_{IN}$ $v_m$/M$_s$Vol.) in nanomagnets. The stars represent switching probability for a 10kT magnet and all other solid lines are for 0kT magnets (E$_A$=0). Note that for the 10kT magnets, an additional E$_A$/C $v_m$ voltage needs to be applied to enable switching in the same manner as a 0kT magnet consistent with the characteristics explained by Eq.(1) and demonstrated in Fig.1.

Additionally, we estimate switching energies and the time associated with the write operation. The results are shown in Fig.4. The switching time of pseudo-magnetization is related to the magnet dynamics. The voltage generated stress can be expressed as an effective magnetic field H$_s$ ≈ (Q$v_m$/M$_s$Vol) ≈ (CV$_{IN}v_m$/MsVol). This effective magnetic field can be used to estimate the typical switching time of magnetization where $\tau \sim 1/\alpha\gamma H_s$ that can result in a few ns switching speeds for typical parameters [34-35], which is similar to typical MRAM structures [33, 36]. Here, α is the Gilbert damping co-efficient and γ is the gyromagnetic ratio. Fig. 4 demonstrates sub-ns switching speeds at room temperature considering α=0.1, which was chosen for faster simulation purposes, for typical values of α~0.01, switching speeds in the few ns range are expected. As the ME heterostructure is a fully capacitive system the write energy approximated by CV$_{IN}^2$/2 can also be very low. Considering a structure with low barrier magnets coupled with an aggressively scaled PE layer and ignoring parasitic and other non-idealities, this number can be optimistically in the ~aJ range for our experimentally guided parameters, which is orders of magnitude smaller than typically reported fJ-pJ write energy ranges of STT-MRAM structures [33,37].

**Experimental Results**

Our proposed memory device is inspired by the recent demonstrations of electrical read-out of the magnetoelectric effect [14-15]. A key parameter that determines the performance of our device is the magnetoelectric back voltage $v_m$ which depends on the stress-induced magnetic field H$_s$ in the heterostructure. Material combinations that will demonstrate large $v_m$ and maintain the $v_m$ after patterning to nanodots are desired. To verify the fundamental device concept and characterize PE/FM structures to extract $v_m$ values, we employ ferromagnetic resonance (FMR) in CoFeB films and nanodots. In the following sections, we introduce how to extract $v_m$ values from our fabricated device structures.

## A. Write Operation in CoFeB Films and Modified Kittel Equation

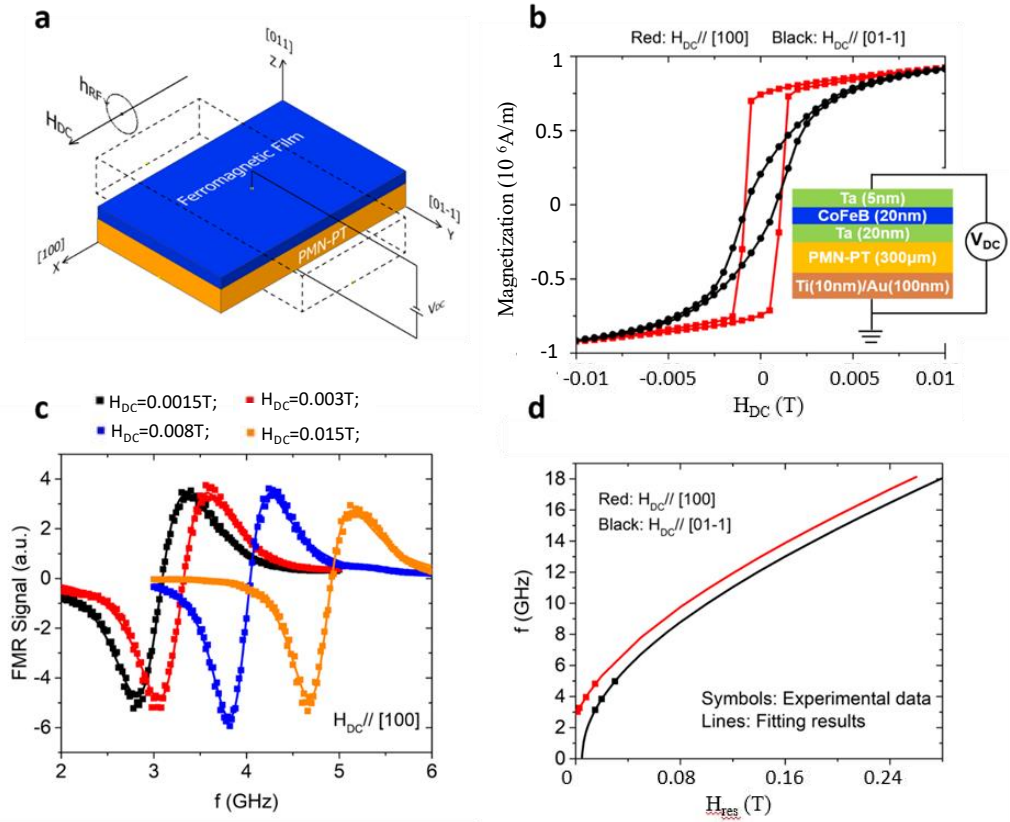

Fig.5. [Experimental data] (a) Configuration of ferromagnetic resonance measurement with $H_{DC}$ applied in the [100] direction when an external DC voltage $V_{DC}$ is applied across the PE/FM heterostructure. The dashed lines show schematically the piezo-response of PMN-PT under a sufficiently positive DC voltage. (b) Magnetic Hysteresis loops M(H) of the stack measured in different directions for $V_{DC} = 0V$ in a SQUID magnetometer. The stack structure of the sample is shown in the inset. (c) FMR spectrum as a function of RF frequencies when $H_{DC}$ is 0.0015T, 0.003T, 0.008T and 0.015T respectively. $H_{DC}$ is applied along the [100] direction. (d) The dependence of RF field frequency and the resonance fields of the sample with $H_{DC}$ applied in the [100] and [01-1] directions with $V_{DC} = 0V$. Lines are fitting results using Kittel equation.

FMR is an effective way to characterize voltage induced magnetization modulation in PE/FM heterostructures [38-50]. Experimentally, we have deposited CoFeB films on (011)-cut PMN-PT substrates and characterized the WRITE operation discussed above using voltage dependent FMR measurements. The stack structure is Ta(5nm)/CoFeB(20nm)/Ta(20nm)/PMN- PT(300μm)/Ti(10nm)/ Au(100nm) as shown in the inset of Fig.5(b). Fig.5(a) shows the experimental set-up of FMR measured along the [100] direction of PMN-PT with an external voltage $V_{DC}$ applied across the PE/FM heterostructure. The dashed lines show that under a sufficiently positive DC voltage, the (011)-cut PMN-PT induces a tensile and a compressive strain in the [01-1] and [100] directions, respectively [14]. Due to mechanical coupling between the PE and FM layers, the biaxial strain in PMN-PT transfers to the CoFeB film and modifies its magnetic anisotropy.

The red and black curves in Fig.5(b) are magnetic hysteresis (MH) loops measured along different directions when $V_{DC}$ = 0V. The square-shaped MH loop and smaller magnetic saturation field in the red curve indicate that the CoFeB film favors the [100] direction as the easy axis. In this case, there exists a tensile strain in the [100] direction and a compressive strain in the [01-1] direction, which is induced by the pre-poled (011)-cut PMN-PT substrate [14]. More details on the piezo-response of PMN-PT under an external DC voltage and the resulting modulation of the CoFeB magnetization properties are described in Ref [14]. Fig.5(c) shows the FMR spectrum, i.e., the first derivative of the absorbed power measured by lock-in technique with a 0.0002T perturbing RF magnetic field $h_{RF}$ applied, as a function of the RF field frequency with different DC magnetic fields applied along the [100] direction when $V_{DC}$ = 0V. The experimental dependence of the resonance fields and the RF signal frequencies are illustrated by the symbols in Fig.5(d). The red and black colors correspond to the scenarios where measurements were performed with $H_{DC}$ applied along the [100] or [01-1] direction, respectively. By fitting the experimental data with the Kittel equation for easy and hard axis [51],

$$f_{FMR\_X} = \frac{\gamma}{2\pi}\sqrt{(H_D + H_k + |H_{res}|)(H_k + |H_{res}|)} \quad (2a)$$

$$f_{FMR\_Y} = \frac{\gamma}{2\pi}\sqrt{(H_D + |H_{res}|)(-H_k + |H_{res}|)} \quad (2b)$$

where $H_D$ is the demagnetizing field for the thin film and $\frac{\gamma}{2\pi}$ is the gyromagnetic ratio, the magnetic anisotropy $H_k$ was extracted to be 0.006T along both the [100] and [01-1] directions.

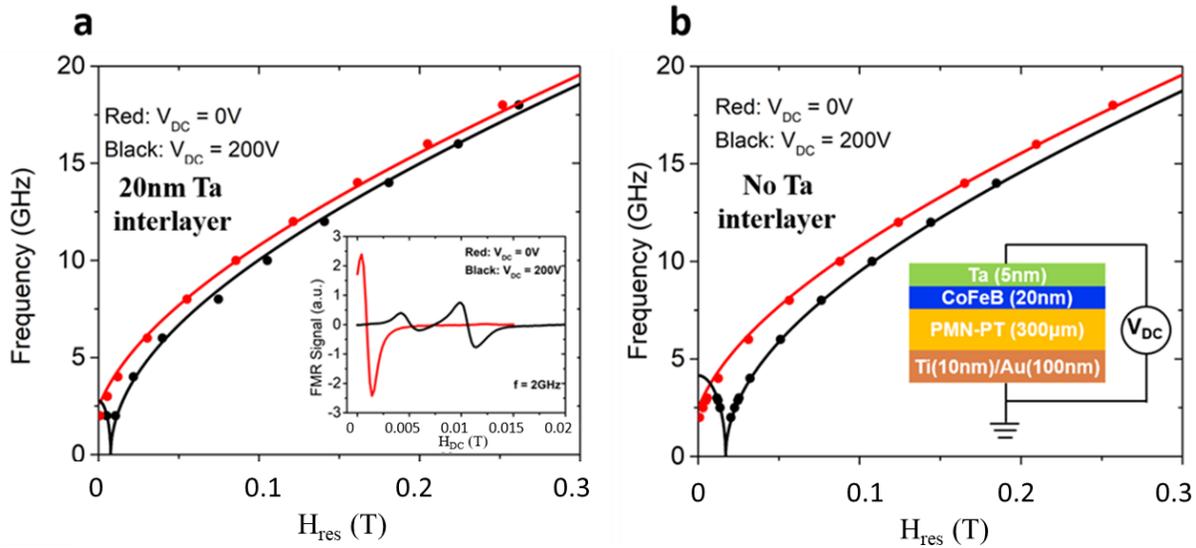

Fig.6. [Experimental data] (a) The dependence of RF field frequency and resonance fields of a Ta(5nm)/CoFeB(20nm)/Ta(20nm)/PMN-PT(300μm)/Ti(10nm)/Au(100nm) sample for $V_{DC}$ = 0V and 200V, respectively. The inset illustrates FMR spectra for $V_{DC}$ = 0V and 200V, respectively when the frequency of the RF field is 2GHz. (b) The dependence of RF field frequency and resonance fields of another sample without a 20nm Ta interlayer between CoFeB and PMN-PT for $V_{DC}$ = 0V and 200V, respectively. The symbols are experimental data, and the solid lines are simulation results calculated by the Eq.(4) and Eq.(5). In both data sets $H_{DC}$ // [100].

As discussed above, when DC voltages are applied, the magnetic anisotropy of CoFeB film is modulated by the strain induced in the PMN-PT film and the resonance frequency of FMR changes accordingly. The inset of Fig.6(a) illustrates the FMR spectra of a Ta(5nm)/CoFeB(20nm)/Ta(20nm)/PMN-PT(300μm)/Ti(10nm)/Au(100nm) sample for $V_{DC} = 0V$ and 200V respectively with $H_{DC}$ applied along the [100] direction when the frequency of the RF field is 2GHz. Compared with the $V_{DC} = 0V$ case, when $V_{DC} = 200V$, the resonance field shifts to the right. This is because when $V_{DC} = 200V$ is applied, compressive strain is induced in the [100] direction and tensile strain is induced in the [01-1] direction due to the re-orientation of the ferroelectric polarization in PMN-PT, which transfers to the CoFeB thin film and provides an in-plane magnetic anisotropic field that makes magnetization along the [100] direction harder. Thus, the resonance field in the [100] direction becomes larger and the FMR curves shift to the right. Note that for measurements at low frequencies along the hard axis (black data points) an additional peak in the FMR signal becomes observable due to a low field solution of the Kittel equation. It is a weak resonance because at low DC magnetic fields the magnetic resonance condition $H_{RF} \perp M$ is not fulfilled [38]. The experimental dependence of RF field frequency and resonance fields of this sample for $V_{DC} = 0V$ and 200V is summarized by the symbols shown in Fig.6(a). Symbols in Fig.6(b) display experimental results obtained on another sample without a 20nm Ta interlayer between CoFeB and PMN-PT, which exhibits a stronger coupling between the piezoelectric and the magnet. To characterize the voltage induced magnetization modulation, some previous reports fitted the experimental results with the Kittel equation and considered the shift of the FMR resonance field as a modulation of the magnetic anisotropy field [52]. However, with an external voltage applied, the Kittel equation must be properly modified to describe the FMR behavior in a PE/FM heterostructure, since a new energy term – the magnetoelastic energy - needs to be included in the calculation. In this case, for an in-plane magnet with an external magnetic field $H_{res}$ applied along its easy (x) axis, the total free energy is,

$$E = M_S Vol. \left\{ \frac{H_k}{2}(1 - m_x^2) + \frac{H_D}{2}m_z^2 - \underbrace{\frac{Qv_m}{M_S Vol.}}_{H_S/2}(m_x^2 - m_y^2) - H_{res}m_x \right\} \quad (3)$$

where $H_D$ is the demagnetization field, $H_S = \frac{2Qv_m}{M_S Vol.}$ is defined as the strain-induced magnetic field, $Q$ is the charge on the PE capacitor, which is proportional with $V_{DC}$, and $v_m$ is the magnetoelectric coefficient. By minimizing the free energy with respect to $H_{res}$, when $H_{res} > -H_K - 2H_S$, $m_x \to 1$; when $H_{res} < -H_K - 2H_S$, $m_x \to -\frac{H_{res}}{2H_S + H_K}$. The positive and negative signs mean the direction of $H_S$ is along the $x$ and $y$ axis, respectively. $H_S \to -\infty$ describes a stress field that is infinite along the $y$-axis. Under this limit, $m_x \to 0$, and the magnetization rotates from $x$ axis to the $y$ axis. When the external magnetic field is applied along the easy axis - [100] direction, the resonance frequency can be derived from the free energy of ferromagnets, and a modified Kittel Equation is obtained as:

$$f_X = \frac{\gamma}{2\pi}\sqrt{(H_D + H_k + |H_{res}| + H_S)(H_k + |H_{res}| + 2H_S)}, \qquad H_{res} > -H_K - 2H_S, \quad (4)$$

$$f_X = \frac{\gamma}{2\pi}\sqrt{\frac{(H_D + H_k + |H_{res}| + H_S)(H_{res}^2 - (H_k + 2H_S)^2)}{H_k + 2H_S}}, \qquad H_{res} \leq -H_K - 2H_S \quad (5)$$

More details on the derivations and comparison between Kittel and the modified Kittel equations are discussed in the Supplementary Information. The solid lines in Fig.6 are fitting results with Eq.(4) and Eq.(5). When $V_{DC} = 0V$, there is no charge on the PE capacitor, thus $H_S = 0$. By fitting the experimental data for this case with Eq.(2), the magnetic anisotropic field $H_k$ and demagnetization field $H_D$ can be obtained as $H_k = 0.006$T and $H_D = 0.8*4\pi M_S$, where $M_S = 1040 \times 10^3$A/m is extracted from the M-H curves for both samples with and without Ta interlayer. With the extracted $H_k$ and $H_D$, by fitting the experimental results for $V_{DC} = 200$V case, the strain-induced magnetic field $H_S$ and magnetoelectric coefficient $v_m$ can be extracted. For the sample with a Ta interlayer, $H_S = -0.004$T and $v_m = 20$mV. For the sample without a Ta interlayer, $H_S = -0.0068$T and $v_m = 34$mV. Note that, this is very close to the $v_m=35$mV extracted value reported in Ref [15]. The extracted $v_m$ -values in Ref [14] are however much lower than the ones reported here, which may have been a result of some signal loss due to the external circuitry used in Ref [14]. The strain induced magnetic field $H_S$ is smaller than the one reported in Ref.[11], we believe this is caused by the non-ideal interface coupling and non-ideal interface quality between CoFeB and PMN-PT as discussed in our previous work [14].

The corresponding effective stability ($\Delta$) of $\mu$ for these two cases – CoFeB film with and without Ta interlayer along with the one reported in Ref [15] are shown in Fig.3(b). Fig.3(b) indicates that the stability of the memory device can be tuned to be in the desired range of 30-80kT Ref [8] for both cases for capacitances in the range of a few hundreds of aF. Note that $v_m$ for the sample with a Ta interlayer is smaller, probably due to the imperfect mechanical coupling between the PE and FM layers [53]. Also note that the two branches of the data at $V_{DC}=200$V in Fig.6(b) (black lines) were fitted using the same set of parameters with Eq.(4) and (5) respectively, which further underlines the validity of the modified Kittel equations.

**B. Write Operation in CoFeB Nanodots**

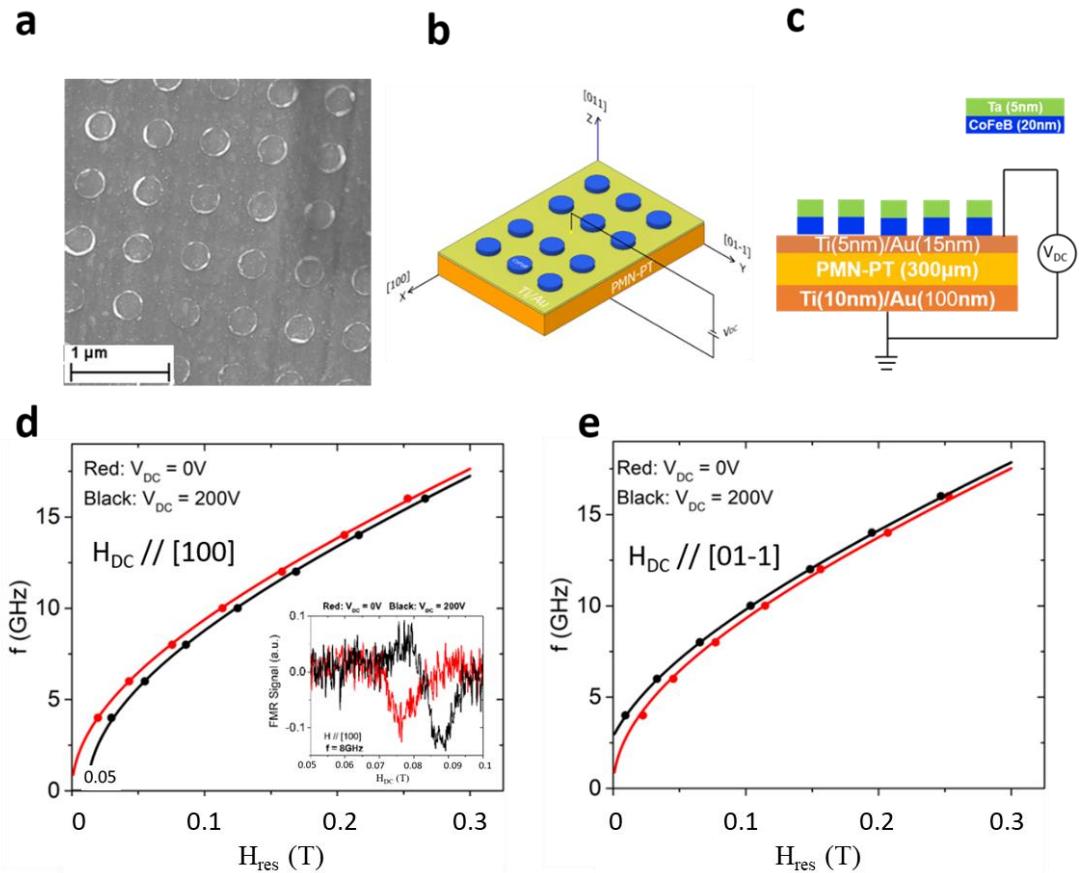

Fig.7. [[Experimental data]]. (a) SEM image of CoFeB nanodots on a PMN-PT substrate. The diameter of the CoFeB dots is 200nm. (b) Schematic of a Ta(5nm)/CoFeB(20nm)/Au(15nm)/Ti(5nm)/PMN-PT(300μm)/Ti(10nm)/ Au(100nm) sample and the measurement configuration. The top 20nm CoFeB with 5nm Ta capping layer were patterned into circular pillars. (c) Material stack of the nanodots sample. (d) The dependence of the RF field frequency and the resonance fields when $V_{DC}$ = 0V and 200V with $H_{DC}$ applied in the [100] direction and (e) the [01-1] direction, respectively. The inset of (d) illustrates the FMR spectra for $V_{DC}$ = 0V and 200V, respectively when the frequency of the RF field is 8GHz as $H_{DC}$ // [100]. The symbols are experimental data and the solid lines are simulation data according to the modified Kittel Equations (4) and (5). Note that (d) and (e) were measured from a nanodots array, the number of nanodots is ~$2.4 \times 10^7$ .

Furthermore, we also fabricated CoFeB nano-magnets on PMN-PT substrates with a structure similar to the one proposed in Fig.1 to verify the WRITE operation. A representative SEM image of 200nm circular nanomagnets is shown in Fig.7(a). Fig.7(b) shows the schematic of the sample structure and the measurement configuration. The material stack of the nanodots sample is shown in Fig.7(c). Fig.7(d) displays the dependence of the RF field frequency and the resonance fields for $V_{DC}$ = 0 and 200V with $H_{DC}$ applied in the [100] direction. The dependence of the FMR signals on the DC magnetic field is illustrated in the inset when the frequency of the RF field is 8GHz. Note that the magnetic anisotropic field of circular nano-magnets $H_k$ ~ 0. We have used a negligible $H_k$ in our model, having a small energy barrier (Δ), through a finite $H_k$ would not appreciably change our conclusions, since for small Δ, magnet behavior is essentially unchanged [25].

Also, when $V_{DC}$ = 0V, $H_S$ = 0. By fitting the experimental results when $V_{DC}$ = 0V with Eq.(2), the demagnetization field $H_D$ was extracted to be $0.65*4\pi M_S$. Next, knowing $H_D$, by fitting the experimental results for $V_{DC}$ = 200V with Eq.(4), $H_S$ and $v_m$ were extracted to be -0.0042T and 21mV respectively. Fig.7(e) shows FMR results of the same sample measured with $H_{DC}$ applied in the [01-1] direction. In this case, when $V_{DC}$ = 0V, the experimental results are the same as the one obtained for the [100] direction. With the same $H_D$ extracted from Fig.7(d), the experimental results were fitted when $V_{DC}$ = 200V with the equation:

$$f_Y = \frac{\gamma}{2\pi}\sqrt{(-H_K + H_{res} + 2H_S)(H_{res} + H_D + H_S)} \quad (6)$$

Eq.(6) is derived (see the Supplementary Information) from the total magnetic free energy term described in Eq.(3) when a magnetic field is applied along the hard (y) axis which is defined here by the PMN-PT crystal. For perfectly circular magnets where $H_k \sim 0$, Eq.(4) and (6) are essentially the same and in the absence of the stress-induced field $H_S$, the equations reduce to the Kittel equations, Eq. (2a,b). $H_S$ and $v_m$ were extracted to be 0.0036T and 18mV respectively which are very close to the numbers extracted for the [100] direction. The small discrepancy in the magnitudes of extracted $H_S$ and $v_m$ along these two directions is likely a result of the fabricated nanodots were not perfectly circular which cause its magnetic anisotropy field is not exactly 0 when $V_{DC}$=0V. Note that there is a Ti(5nm)/Au(15nm) interlayer between the PMN-PT substrate and CoFeB nano-magnets and the extracted $H_S$ and $v_m$ quantities are very similar to the ones obtained in CoFeB films when there is a 20nm Ta interlayer between the PE and FM layers which further supports the validity of our modified Kittel formula. Note that Fig.3(b) had indicated that the coupled capacitance value can be adjusted to get the desired stability for any $v_m$ value, but in reality, we would not want the back-voltage $v_m$ value to be too small as this is the voltage that facilitates the reading operation. Our results indicate that even after patterning and having an interlayer between the CoFeB/PMN-PT stack, the heterostructure was able to retain large enough $v_m$ to facilitate READ-WRITE operation for our envisioned memory device.

## Conclusion

In conclusion, we have proposed a novel magnetoelectric memory device based on a pseudo-magnetization in a PE/FM heterostructure. The use of low-barrier nano-magnets as the FM layer and read-out through the magnetoelectric back-voltage can potentially enable the device to operate with an energy-delay of hundreds of aJ to ps, combining attractive features of magnetic and ferroelectric memory technologies, such as high-density and non-volatility. The emergent ferroelectricity from an FM/PE heterostructure will lead to interesting new physics. The experimental results indicate that FMR is an effective way to characterize the magnetic WRITE operation. To quantify the strain induced magnetic field, we have derived a modified Kittel formula by including a magnetoelastic energy in the magnetic free energy. According to our analysis, the magnetoelastic coefficient $v_m$ between CoFeB and PMN-PT is in the tens of mV range which is significant for proper device operation, since it is the output voltage and has to be above the noise level. We believe that $v_m >20$mV is a desirable target, since stable states can be achieved with the use of capacitors in the hundreds of aF range, consistent with typical capacitance values in modern 14nm transistors. Furthermore, according to our results, the presence of an interlayer between the PE and FM layer does not degrade the $v_m$ value too much and pattering also did not degrade $v_m$. The theoretical and experimental work presented here indicates that the proposed novel ME memory device is feasible and relevant for the development of future energy efficient spintronics devices.

## Methods

CoFeB nano-magnets were fabricated on one side of the poled (011)-cut PMN-PT substrate using e-beam lithography, metal deposition and a conventional lift-off process. Both CoFeB nano-magnets and CoFeB films were deposited in a multisource magnetron sputtering system with a base pressure of $3*10^{-8}$Torr followed by a 5nm Ta deposition in the same system. Ti/Au layers were deposited in an e-beam evaporation system. Ferromagnetic resonance measurements were performed using a NanOsc CryoFMR probe in a Quantum Design DynaCool PPMS. It is a coplanar waveguide type of broadband FMR spectrometer. Magnetization hysteresis loop measurements were performed in a Quantum Design MPMS-3 SQUID magnetometer.

**Supplementary Information:** Derivation of the modified Kittel Equation used in text.

## Author's Contributions

T.S. and O.H. contributed equally to this work. T.S worked on the device fabrication, characterization, and data analysis; O.H. and K.Y.C. worked on the theoretical formulations and simulations; S.D. worked on the theoretical formulations, N.D. worked on the experimental measurements; J.A. analyzed the data and oversaw the planning and execution of the project; T.S. and O.H. wrote the manuscript.

## Acknowledgements

We gratefully acknowledge useful discussions with Sunil A. Bhave. This work was supported in part by the Center for Probabilistic Spin Logic for Low-Energy Boolean and Non-Boolean Computing (CAPSL), one of the Nanoelectronic Computing Research (nCORE) Centers as task 2759.003 and 2759.004, a Semiconductor Research Corporation (SRC) program sponsored by the NSF through CCF 1739635. The authors thank the staff at the Birck Nanotechnology center for their technical support.

## Data Availability statement

The data that support the findings of this study are available from the corresponding author upon reasonable request.

**Competing interests**

The authors declare no competing interest.